\documentclass{WileyMSP-template}

\usepackage{graphicx}
\usepackage{bm}
\usepackage{amsfonts}
\usepackage{amssymb}
\usepackage{amsmath}
\usepackage[colorlinks=true,linkcolor=blue]{hyperref}%
\usepackage{color}

\DeclareRobustCommand{\_}[1]{\sb{\mathrm{#1}}}

\begin{document}
\title{Direction-sensitive magnetophotonic surface crystal} 
\maketitle

\author{Richard M.\ Rowan-Robinson*}
\author{J{\'e}rome Hurst}
\author{Agne Ciuciulkaite}
\author{Ioan-Augustin Chioar}
\author{Merlin Pohlit}
\author{Mario Zapata}
\author{Paolo Vavassori}
\author{Alexandre Dmitriev*}
\author{Peter M.\ Oppeneer}
\author{Vassilios Kapaklis*}

\begin{affiliations}
R. M.\ Rowan-Robinson \\
Department of Material Science and Engineering, University of Sheffield, Sheffield, United Kingdom \\
Email: r.rowan-robinson@sheffield.ac.uk \\
\vspace{4pt}
J. Hurst \\
Univ. Grenoble Alpes, CNRS, CEA, Grenoble INP, IRIG-Spintec, F-38000 Grenoble, France \\
\vspace{4pt}
R. M.\ Rowan-Robinson, J. Hurst, A. Ciuciulkaite, I-A. Chioar, M. Pohlit, Prof. P. M.\ Oppeneer, Prof. V. Kapaklis\\
Department of Physics and Astronomy, Uppsala University, Box 516, SE-751 20 Uppsala, Sweden \\
Email: vassilios.kapaklis@physics.uu.se \\
\vspace{4pt}
I-A. Chioar \\
Department of Applied Physics, Yale University, New Haven 06511, CT, USA \\
\vspace{4pt}
M. Zapata, Prof. P. Vavassori \\
CIC nanoGUNE BRTA, E-20018 Donostia-San Sebastian, Spain \\
\vspace{4pt}
M. Zapata, Prof. P. Vavassori \\
IKERBASQUE, Basque Foundation for Science, E-48013 Bilbao, Spain \\
\vspace{4pt}
Prof. A. Dmitriev \\
Department of Physics, University of Gothenburg, SE-412 96 G\"oteborg, Sweden \\
Email: alexd@physics.gu.se \\
\end{affiliations}

\keywords{magnetoplasmonics, rare-earth--transition-metal ferrimagnets, all-optical switching, magnetophotonic crystal, Fano resonance}

\begin{abstract}
Nanometer-thin rare-earth-transition metal (RE-TM) alloys with precisely controlled compositions and out-of-plane magnetic anisotropy are currently in the focus for ultrafast magnetophotonic applications. However, achieving lateral nanoscale dimensions, crucial for potential device downscaling, while maintaining  designable optomagnetic functionality and out-of-plane magnetic anisotropy is extremely challenging. Here we integrate nanosized Tb$_{18}$Co$_{82}$ ferrimagnetic alloys, having strong out-of-plane magnetic anisotropy, within a gold plasmonic nanoantenna array to design micrometer-scale a magnetophotonic crystal that exhibit abrupt and narrow magneto-optical spectral features that are both magnetic field and light incidence direction controlled. The narrow Fano-type resonance arises through the interference of the individual nanoantenna's surface plasmons and a Rayleigh anomaly of the whole nanoantenna array, in both optical and magneto-optical spectra, which we demonstrate and explain using Maxwell-theory simulations. This robust magnetophotonic crystal opens the way for conceptually new high-resolution light incidence direction sensors, as well as for building blocks for plasmon-assisted all-optical magnetization switching in ferrimagnetic RE-TM alloys.\\
\end{abstract}

\section{Introduction}
Nanoscale magnetophotonics merges magnetism with nanophotonics \cite{maccaferri2020}, combining seamlessly magneto-optical (MO) effects with surface plasmons, thus being capable of delivering ultra-high performance biological and chemical sensors \cite{maccaferri2015a, zubritskaya2015}, active tunability in nano-optics by external magnetic fields \cite{maccaferri2020, temnov2010, zhang2017, torrado2010, belotelov2007, belotelov2011, gonzalez-diaz2007, rowan-robinson2019, rollinger2016, lodewijks2014}, and setting a platform for ultrafast opto-magnetism and spintronic \cite{liu2015a} devices on the nanoscale. Pure ferromagnetic plasmonic systems were earlier considered unfeasible for these purposes due to the high ohmic losses associated with the transition-metal ferromagnets. However, to a large extent, these can be overcome through nanopatterning \cite{ctistis2009, papaioannou2010}, materials engineering and fabrication of hybrid noble metal-ferromagnetic nanostructures \cite{zubritskaya2018, kataja2016, martin-becerra2010, pourjamal2018, banthi2012}. The enhancement of various MO effects is typically achieved in these systems through near-field light concentration at the nanoscale, boosting light-magnetism interactions that relate to the MO Voigt parameter of the ferromagnet \cite{moncada-villa2014, gonzalez-diaz2008}. Importantly, by exploiting magnetic anisotropy control, the magnetization can be stabilized in a desired direction and MO effects can be recorded at zero external magnetic field. Linewidth engineering \cite{kataja2016, kataja2015, maccaferri2016} wherein high Q-factor resonances are achieved, can furthermore be employed in ordered arrays of magnetoplasmonic nanoantennas with surface lattice resonances. 

The use of rare-earth--transition-metal alloys is of paramount interest for future nanoscale magnetophotonic and magnetoplasmonic systems for several key reasons. Firstly, they are known to exhibit very large MO effects \cite{buschow1989, atkinson1988} potentially permitting very high real-time active tunability of light polarization. Secondly, they can exhibit strong perpendicular magnetic anisotropy, yet with an amorphous texture \cite{ciuciulkaite2020, yoshino1984, hebler2016, harris1992, frisk2016a}. For instance, carefully engineered Co/Pt multilayered nanodots having large interfacial spin-orbit coupling with perpendicular magnetic anisotropy, demonstrate tenfold enhancements in MO activity and demonstrate the great potential of out-of-plane magnetic anisotropy materials for magneto\-plasmonics \cite{freire-fernandez2020}. The amorphous texture of RE-TM alloys greatly simplifies the otherwise stringent requirements on material microstructure for obtaining these highly desired magnetic properties. As such, they can be grown on noble metals like Au with minimal residual stresses, and with highly smooth interfaces, thereby maintaining much of their original magnetic properties even after patterning \cite{ciuciulkaite2020}. Importantly, with perpendicular magnetic anisotropy the remanent magnetization state of the magnetic nanostructures can be designed to be parallel to the light propagation direction for normal light incidence, greatly simplifying potential practical applications of magnetoplasmonic crystals. This allows one to explore their MO functionality (such as, tunable Faraday effect) directly, i.e., without the need of external magnetic fields in order to stabilize the magnetization along the out-of-plane axis. Thirdly, ferrimagnetic alloys such as Tb$_{18}$Co$_{82}$, have recently experienced extensive interest due to the demonstration of enhanced spin-orbit torques \cite{finley2016, ueda2016, alebrand2012} and all-optical switching \cite{alebrand2012, mangin2014,AvilesFelix2020,ciuciulkaite2020}, allowing for zero-field magnetic switching, on picosecond timescales, with the use of pulsed lasers. Thus, demonstrating the compatibility of these materials with nanoantennas is essential for the development of nanoscale (i.e., sub-diffraction) all-optical switching technologies \cite{liu2015a}.

Here we devise a magnetophotonic crystal composed of nanocone Au plasmonic nanoantenna arrays incorporating an amorphous RE-TM ferrimagnetic alloy,
Tb$_{18}$Co$_{82}$, with perpendicular magnetic anisotropy \cite{ciuciulkaite2020}. We show that this hybrid Au/Tb$\_{18}$Co$\_{82}$ system provides high-Q MO resonances, overcoming the losses associated with ferrimagnetic alloys. By Maxwell-theory modelling we show that this is achieved through the resonant collective excitation of surface lattice modes that exhibit a particularly strong angular dispersion. This is a result of the interference of a Rayleigh anomaly with the individual nanoantennas' plasmons, giving rise to surface lattice resonance resonances with characteristic Fano-type asymmetric lineshape in both the optical and MO spectra. We demonstrate an exceptionally strong tunability of the spectral position of such resonances by varying the angle of incidence (incident direction) of the incoming light, exemplifying the potential of magnetophotonic crystals for high-resolution mechanical tilt-angle sensors and, more broadly, for actively-controlled optical systems \cite{haghtalab2020,shi2020}.

\section{Results and Discussion}

\begin{figure}[t!]
\centering
     \includegraphics[width = 0.7\textwidth]{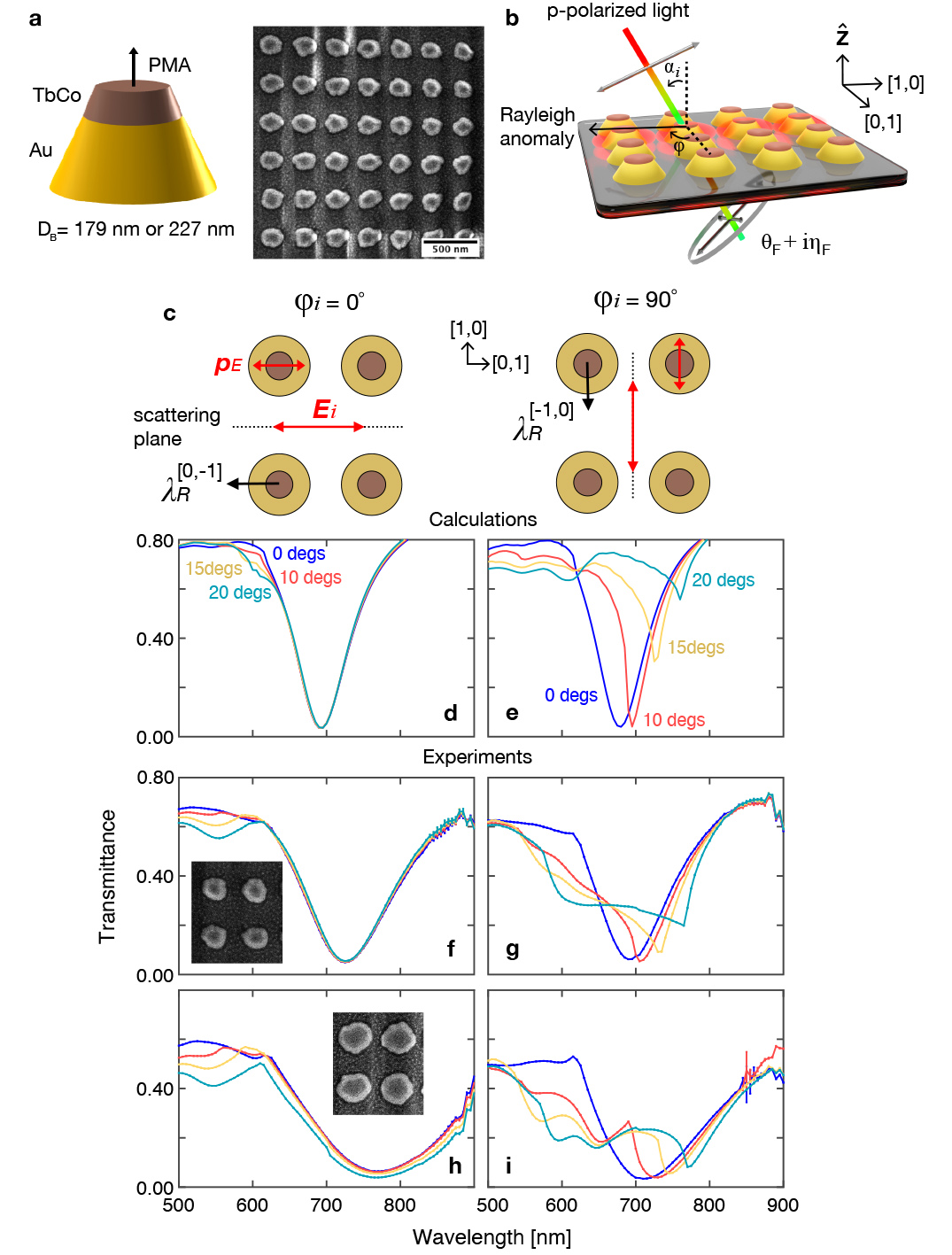}
  \caption{ Magnetophotonic crystals composed of arrays of truncated nanocone hybrid antennas, with tunable optical transmission response. (a) Schematic of a single Au-TbCo nanoantenna featuring PMA (left) and scanning electron micrograph view of a magnetophotonic crystal (right). (b) Magnetophotonic crystal illumination with resulting Faraday rotation ($\theta\_{F}$), ellipticity ($\eta\_{F}$) of the transmitted light, and the Rayleigh anomaly associated with the passing-off of the diffraction order. (c) Magnetophotonic crystal illumination with two azimuthal orientations ($\varphi\_{i}$ = 0 and 90$^\circ$) with respect to the incident light polarisation ($E\_i$) and scattering plane, with $p\_E$ denoting the orientation of the electric dipolar plasmon in the nanoantennas. The reciprocal lattice vectors $[1, 0]$ and $[0, 1]$ are shown to illustrate the 90$^{\circ}$ rotation of the reciprocal lattice vectors with respect the real space lattice. (d, e) Calculated transmission spectra for incidence angles $\alpha\_{i}$ between 0 and 20 degrees, for the $\varphi\_i$ = 0 (d) and the $\varphi\_i$ = 90$^\circ$ (e) configurations, respectively. (f, g) Measured transmission spectra for incidence $\alpha\_{i}$ angles 0 - 20 degrees for the magnetophotonic crystal built on  $D\_B$ = 179~nm nanoantennas for the $\varphi\_{i}$ = 0 (f, inset – SEM of nanoantennas in this magnetophotonic crystal) and the $\varphi\_{i}$ = 90$^\circ$ (g). (h, i) Same as (f, g) but for the magnetophotonic crystal with  $D\_B$ = 227~nm nanoantennas (inset in (h) – nanoantennas SEM).
  }
\label{fig1}
\end{figure}

Nanocone antennas were previously shown to exhibit a very strong field enhancement \cite{horrer2013}, with the electromagnetic field concentrated at the tip \cite{horrer2013, schafer2013}. We build large rectangular lattice arrays of Au/Tb$\_{18}$Co$\_{82}$ truncated nanocone antennas (Fig. \ref{fig1}a) \cite{ciuciulkaite2020} with two selected base diameters (179 $\pm$ 5~nm and 227 $\pm$ 4~nm (see SEM insets in Fig. \ref{fig1}f and h respectively). The light incidence angle ($\alpha\_{i}$) is varied with respect to the lattice plane, directed  along  either  one  or the other of the array periodicity (Fig. \ref{fig1}b). We first use finite-element Maxwell-theory simulations (COMSOL  Multiphysics, see Supporting Information) to pinpoint the emerging 
resonances' linewidth narrowing and high incidence-angle sensitivity. The magnetophotonic crytstal is built of Au(80 nm)/Tb$\_{18}$Co$\_{82}$(15 nm) truncated nanocones (base diameter, $D\_{B}$ = 179~nm), arranged in a rectangular array with 340~nm $\times$ 425~nm periodicity (Fig.\ \ref{fig1}c). The light incidence direction angle (using the optical convention) defines a scattering plane which is parallel to one (340~nm) or the other (425~nm) of the array periodicity axes with azimuthal angles $\varphi\_{i}$ = 0 or $\varphi\_{i}$ = 90$^{\circ}$, respectively. 

\newpage

Surface lattice resonances are the result of the coupling between a broad lossy resonance, in this case the localised plasmon resonances of individual nanoantennas, and diffracted waves in the plane of the nanoantenna array (a detailed description is provided in the supporting information). This condition is generally observed close to a Rayleigh anomaly, where for a given $\alpha\_{i}$ and lattice periodicity, a Rayleigh anomaly exists where a diffracted wave is directed parallel to the grating \cite{rayleigh1907}. This Rayleigh anomaly represents the passing-off of a diffraction order through a laterally excited beam. There can exist a large number of these diffraction orders, which are labeled by two integers $n$ and $m$. The allowed waves are obtained by imposing the condition that the component of the light wave-vector normal to the lattice surface is real, through the expression
\begin{equation}
    k_{\perp} = \sqrt{k_s^2 - \left(\bm{k_{\parallel}} + m \bm{G_1} + n \bm{G_2} \right)^2} \in \Re.
    \label{lattice diffraction modes}
\end{equation}
In the above formula, $k\_s = 2\pi n\_{sub} / \lambda $ corresponds to the light wave-vector in the substrate, where $n\_{sub}$ is the refractive index of the fused silica substrate ($n\_{sub}$ = 1.45), $\lambda$ the light wavelength, \newline $\bm{k_{\parallel}}=k_0\left[ \sin(\alpha_i)\cos(\varphi_i)~\bm{u_x} + \sin(\alpha_i)\sin(\varphi_i)~\bm{u_y} \right]$ corresponds to the wave-vector component of the incident radiation (in air/vacuum) parallel to the lattice surface, $k_0 = 2\pi/\lambda$ is the light wave-vector in air and $\bm{G_1} = (2\pi/a) ~\bm{u_x}$, $\bm{G_2} = (2\pi/b) ~\bm{u_y}$ are the reciprocal lattice vectors, with $~\bm{u_x}$, $~\bm{u_y}$ being the reciprocal lattice unit vectors and $a$ = 340 nm, $b$ = 425 nm, being the lattice parameters. The number of diffracted waves depends on the lattice parameters, the angle of incidence, the refractive index of the substrate and the light wavelength. For wavelengths greater than 600 nm, Equation \eqref{lattice diffraction modes} indicates that only the diffracted waves ($n=0 ~;~  m=-1$) for $\varphi_i=0$ and ($n=-1 ~;~  m=0$) for $\varphi_i= 90^{\circ}$ can be obtained by varying the incidence angle between 0 - 20$^{\circ}$ (see Supporting Information). 

We use reciprocal vector notation, such that the Rayleigh anomaly occurs at wavelengths $\lambda\_{R}^{[n,m]}$  with wave-vector orientated along the reciprocal lattice vectors $[n, m]$. Fig.~\ref{fig1}c demonstrates how the reciprocal lattice vectors are orientated with respect to the real-space lattice. The analytical expressions for the two allowed substrate waves ($[0,-1], [-1,0]$) from Equation \eqref{lattice diffraction modes}, are given by
\begin{equation}
\lambda\_{R}^{[0,-1]} = a \left[ n\_{sub}  +  n\_{air} \sin \left( \alpha\_i \right)\right]~~ \textrm{for} ~\varphi\_i = 0,
\end{equation}
\begin{equation}
\lambda\_{R}^{[-1,0]} = b \left[ n\_{sub}  +  n\_{air} \sin \left( \alpha\_i \right)\right]~~ \textrm{for} ~\varphi\_i = 90^{\circ},
\end{equation}

\noindent
where $n\_{air} = 1$ is the refractive index of air.

We first calculate the spectral transmission  through the array for $p$-polarised light (i.e., incident electric field is in the scattering plane) (Fig.\ \ref{fig1}d, e). Individual nanoantenna dipole-type plasmons are excited in the respective scattering planes at 690~nm at normal incidence ($\alpha\_{i}$ = 0). For the $\varphi\_{i}$ = 0 configuration (scattering plane along 340~nm array periodicity, Fig.\ \ref{fig1}d) the surface lattice resonances from Eq. (2) are at $\lambda\_{R}^{[0,-1]}$ = 493~nm, 552~nm, 581~nm and 609~nm for $\alpha\_{i}$ = 0, 10, 15, and 20 degrees, respectively, and therefore not spectrally overlapping with the nanoantennas' individual plasmons. For $\varphi\_{i}$ = 90$^{\circ}$ (scattering plane along 425~nm array periodicity, Fig.\ \ref{fig1}e),  Eq.\ (3) gives $\lambda\_{R}^{[-1,0]}$ = 616 nm, 690 nm, 726 nm and 762 nm, strongly overlapping with the nanoantennas' plasmon, resulting in a very substantial tuning of the spectrally abrupt transmission spectrum by changing $\alpha\_{i}$ (see Fig.\ \ref{fig1}e).

In the Fano-type resonance description \cite{fano1961, lukyanchuk2010}, the nanoantennas plasmon represents a continuum of states, whereas the Rayleigh anomaly is a narrow line-width diffracted wave, which, upon interfering with the continuum, results in the characteristic asymmetric lineshape of the surface lattice resonances. A similar behaviour has been seen previously with magnetoplasmonic Ni nanoantennas arrays \cite{maccaferri2016, kataja2015}, where the overlap between $\lambda\_{R}^{[n,m]}$ and the nanoantenna plasmon was tuned by varying the lattice periodicity of the magnetoplasmonic crystal. However, a much simpler alternative method of tuning the surface lattice resonance spectral position can be obtained using the angular dispersion of $\lambda\_{R}^{[n,m]}$. This tuning of the spectral position of the surface lattice resonance opens up applications as mechanical tilt-angle  transducers/sensors, and in contrast with previously observed transmission/reflectance angular dependence in pure plasmonic arrays \cite{vecchi2009}, this magnetoplasmonic crystal allows one to fully explore angular MO tunability.

The dipolar radiation field is strongest transverse to the dipolar plasmon oscillation given by $p\_{E}$ (Fig. \ref{fig1}c). In our simulations we used $p$-polarised light and hence the electric dipole excitation within individual nanocone antennas is orientated within the scattering plane and parallel to the diffraction anomaly. This dipole cannot radiate along the oscillation direction, hence there must exist an additional mechanism for light to be scattered along the other periodicity direction for the excitation of the Rayleigh anomaly. We show this to be the result of an out-of-plane  component  to  the  electric  dipole due to the illumination at oblique incidence (see  Supporting Information), which would radiate in all directions within the plane of the lattice \cite{huttunen2016} providing the excitation for all $\lambda\_{R}^{[n,m]}$, e.g. [-1,~0], [0,~-1], [-1,~-1] waves for $p$-polarised light.  

The measured transmission spectra are shown in Fig.\ \ref{fig1}f-i. In agreement with the electromagnetic simulations above,  for the $\varphi_{i}$ = 0 configuration (Fig.\ \ref{fig1}f, h) the transmission spectra show very little dependence on $\alpha\_{i}$.  The nanoantenna plasmon is red-shifted and spectrally broadened as compared to the simulations though, which is likely a result of the thin Al$_{2}$O$_{3}$ isolation layer (see Methods) and oxidation of the exposed Tb$_{18}$Co$_{82}$ side-walls on the fabricated nanocones and also the size and shape distribution of the nanoantenna ensemble.  There is a spectral feature between 500 - 600~nm (Fig.\ \ref{fig1}f, h) that migrates to longer wavelengths as $\alpha\_{i}$ increases that is most likely due to $\lambda\_{R}^{[0,-1]}$,  since it occurs at the same spectral positions for both the $D\_{B}$ = 179~nm (Fig.\ \ref{fig1}f) and 227~nm (Fig.\ \ref{fig1}h) nanoantennas, suggesting its origin relates to the lattice and not the individual nanoantenna plasmon resonance.

When rotated into  the $\varphi\_{i}$  = 90$^{\circ}$  configuration (Fig.\   \ref{fig1}g and  i), the strong variations  in  the  transmission spectra  are  observed, in excellent agreement with the simulations, in  both  spectral position and lineshape, albeit  with  reduced amplitude. For both $D\_B$ = 179~nm (Fig.\ \ref{fig1}g) and 227~nm (Fig.\ \ref{fig1}i) nanoantennas, the $\alpha\_{i}$ = 0 incidence shows a small blue shift of the plasmon for the $\varphi_{i}$ = 90$^{\circ}$ configuration relative to the $\varphi_{i}$ = 0 configuration. As shown in the inset scanning electron microscopy images, the nanocones are not perfectly circular and this discrepancy is likely a result of this asymmetry. Markedly, the broad spectral distribution with the $D\_B$ = 227~nm nanocone antennas allows for a larger tuning bandwidth, such that there exists a larger range of $\alpha\_{i}$ for which $\lambda\_{R}^{[-1, 0]}$ overlaps with the nanoantenna plasmon. 

\begin{figure}[t!]
\centering
     \includegraphics[width = 0.85\textwidth]{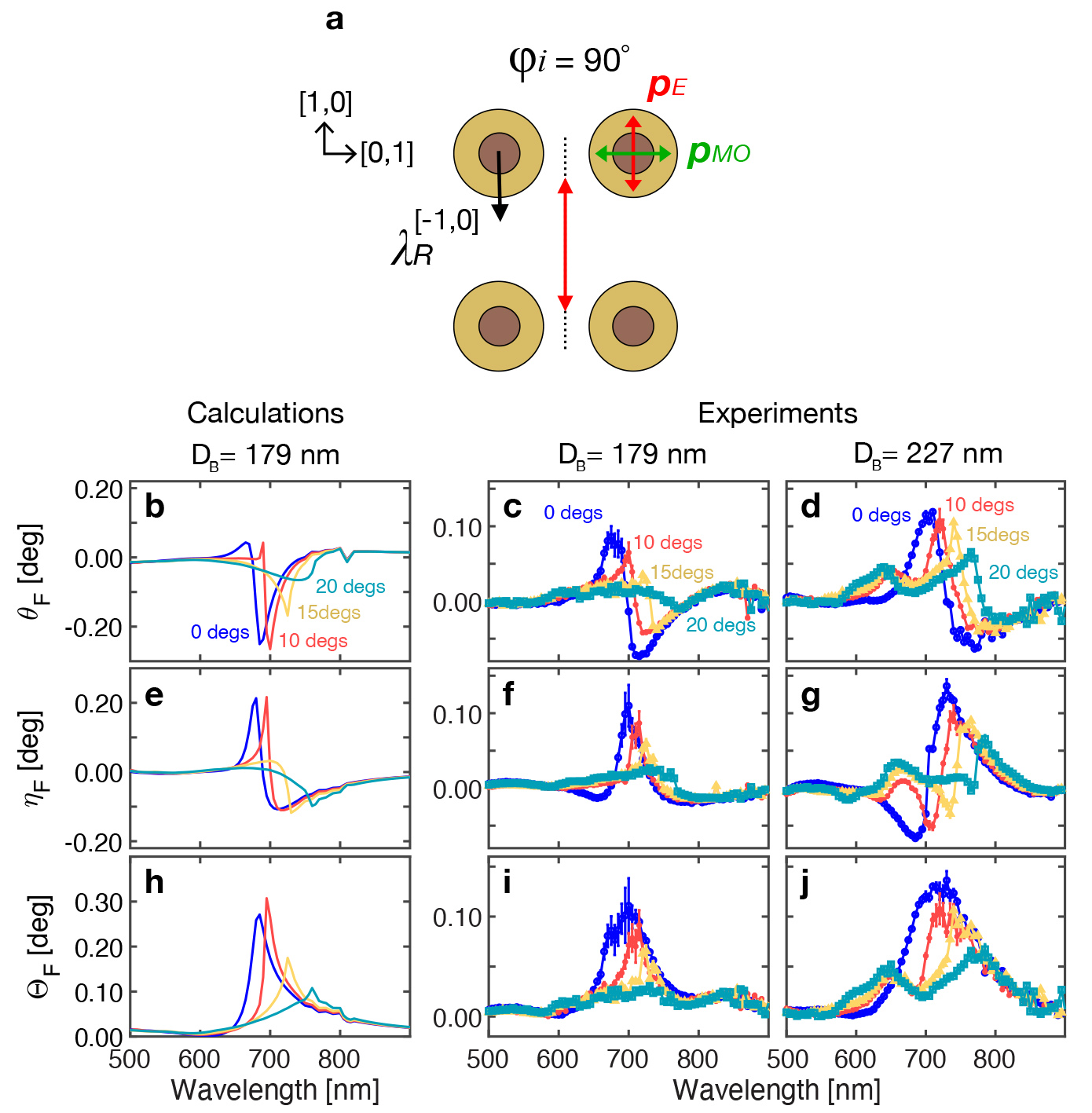}
  \caption{Angle-of-incidence spectral dependence of the Faraday  effect in the magnetophotonic crystals. (a) Illumination configuration as in Fig, 1c, with added MO plasmon dipole of nanoantenna ($p\_{MO}$, green). (b, e, h) Calculated spectral $\theta\_{F}$ (b), $\eta\_{F}$ (e) and $\Theta\_{F}$ (h) for incidence angles $\alpha\_{i}$ between 0 and 20 degrees. Measured $\theta\_{F}$ (c, d), $\eta\_{F}$ (f, g) and $\Theta\_{F}$(i, j) for the $D\_{B}$ = 179 nm and 227 nm nanoantenna arrays respectively. A quadratic polynomial has been fitted to the $\theta\_{F}$  measurements and subtracted to remove the background contribution which arises from the Faraday rotation of the fused-silica substrate, which is strongest for short wavelengths and approaches zero with increasing wavelength.
  }
\label{fig2}
\end{figure}

While we readily earn high incidence direction tunability of optical transmission with the designed magnetophotonic crystals, resonances in MO spectra can yield much larger Q-factors \cite{qin2017}. Maccaferri et~al. \cite{maccaferri2013} showed that an out-of-plane magnetization in the presence of the electric dipolar plasmon gives rise to an in-plane MO dipolar plasmon ($p\_{MO}$) which is orientated orthogonal to $p\_{E}$ and is induced in the ferromagnetic layer. The magnitude of $p\_{MO}$ is proportional to the magnitude of $p\_{E}$. Given that a material's optical constants are typically much larger than their MO constants, even lossy broad localised plasmon resonances can give rise to large enhancements in MO activity as compared to ferrimagnets without plasmonic integration. This transverse oscillation is induced via spin-orbit coupling, generating an oscillation of conduction electrons in-the-plane but orthogonal to $p\_{E}$. With the use of p-polarised light, the pure optical dipole is orientated along $p\_{E}$ and the transverse MO dipole is aligned along $p\_{MO}$ (Fig. \ref{fig2}a).  Hence, the use  of $p$-polarised light results in the MO dipole induced in the Tb$_{18}$Co$_{82}$ layer which radiates strongly in the scattering plane, and is therefore expected to be most sensitive to the angular dispersion of the surface lattice resonances as the crystal is tilted by $\alpha\_{i}$.

In Fig. \ref{fig2}b-j the calculated and experimental Faraday rotation  ($\theta\_F$), Faraday ellipticity ($\eta\_F$) and Faraday angle ($\Theta\_F = \sqrt{\theta\_{F}^{2} + \eta\_{F}^{2}}$) are  presented. The calculated Faraday effect  using the experimental permittivity for a Tb$_{18}$Co$_{82}$ thin film is shown in Fig.\ \ref{fig2}b, e, h for the $D\_{B}$ = 179~nm nanocone antennas array (see Supporting Information for details). The $\varphi\_{i}$ = 0 configuration shows no angular dependence for the Faraday effect (see Supporting information) and through fitting a Lorentzian to the $\alpha\_{i}$ = 0 transmission and $\Theta\_{F}$ spectra for the $\varphi\_{i}$ = 0 configuration we estimate that the MO resonance exhibits a two-fold reduction in linewidth relative to the pure optical resonance. While for $\varphi\_{i}$ = 0 (no overlap of nanoantennas plasmon with Rayleigh anomaly) a reasonable spectral feature narrowing is achieved without angular dependence, in the $\varphi\_{i}$ = 90$^\circ$ configuration the experimental Faraday spectra show strong angular dependence and suggest that sizeable Faraday angles of up to 0.3$^\circ$ are readily available. The simulated spectra prompts that extremely sharp features exist that coincide with $\lambda\_{R}^{[-1,0]}$ (Fig.\ \ref{fig2}b, e, h).

The Rayleigh anomaly is strongest through the substrate and the observation of strong diffractive effects in the Faraday spectra indicates that the MO dipole induced in the Tb$_{18}$Co$_{82}$  layer is transferred to the rest of the nanoantenna \cite{pourjamal2018}. The experimental MO spectra measured for the nanoantennas with $D_{B}$  = 179 nm (Fig.\ \ref{fig2}c, f, i) compare well to the calculations. The excellent match of the measured spectra with simulations  demonstrates the suitability of  combining  finite-element methods with experimentally measured thin-film permittivity for the calculated design of magnetophotonic devices.  For the nanoantennas with $D_{B}$ = 227 nm (Fig.\  \ref{fig2}d, g, j) there is a stronger Faraday effect, but with broader spectral features, demonstrating the trade-off between adding more magnetic material in the nanoantenna whilst maintaining small dimensions for narrow plasmon resonances.

From the above it is clear that it is not possible to measure the MO response of the Au-Tb$_{18}$Co$_{82}$ nanoantennas off-resonance, where $\theta\_F$ and $\eta\_F$ quickly drop to values comparable to the measurement uncertainty. In effect, the nanoantennas plasmons strongly amplify the minute magnetic signals that ordinarily wouldn't be resolved. It is possible to estimate the Tb$_{18}$Co$_{82}$ amount in each nanoantenna, corresponding to a nanodisk with 86 $\pm$ 10~nm diameter and 15~nm height for nanoantennas with base diameter of 179~nm. This yields Tb$_{18}$Co$_{82}$ effective film thickness (i.e.\ the thickness of a film made with the same amount of material) of approximately 0.6~nm, of the order of an atomic monolayer, demonstrating the MO amplification obtained through the nanoantenna's plasmons.

\begin{figure}[t!]
\centering
     \includegraphics[width = 0.85 \textwidth]{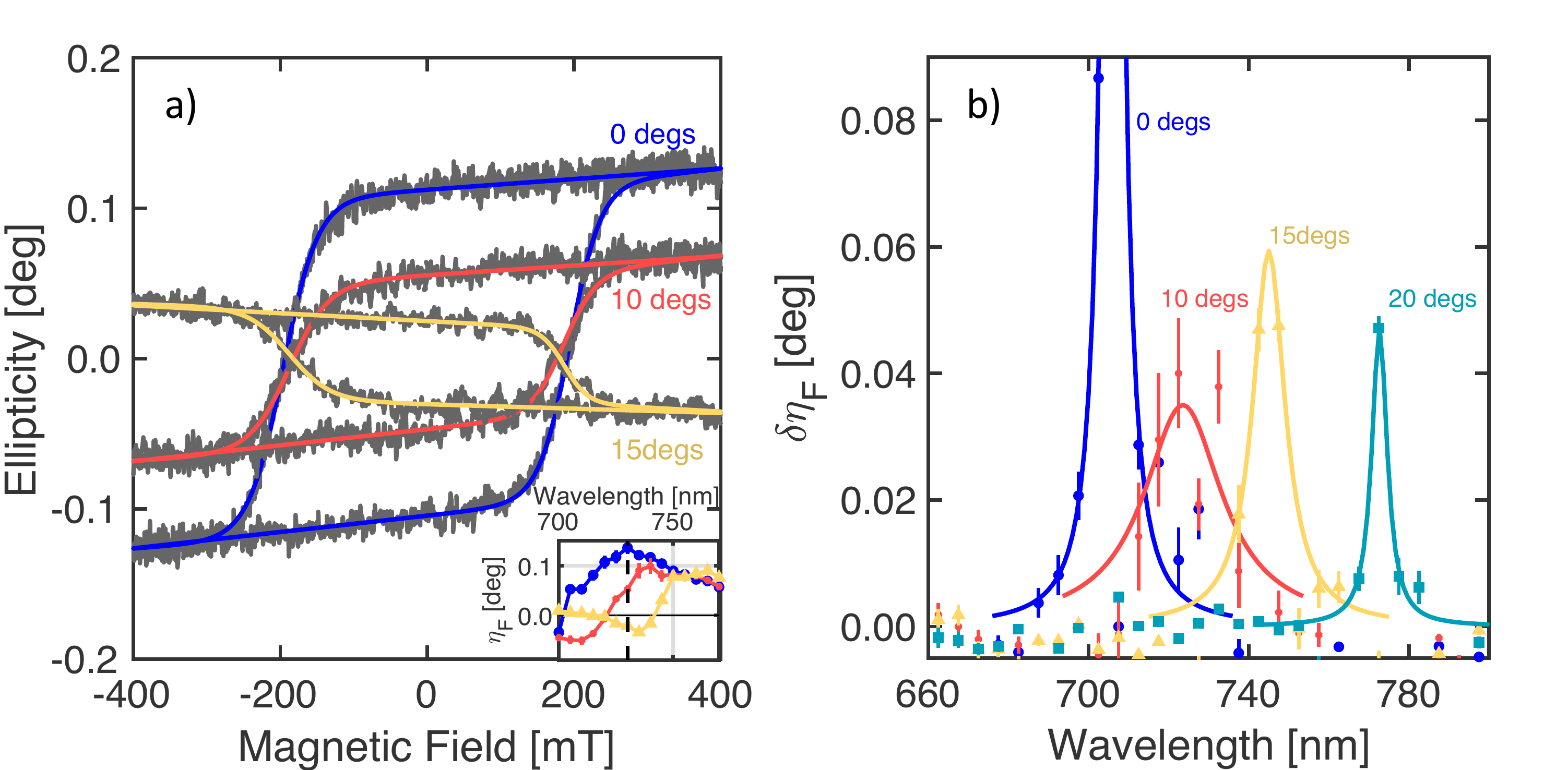}
  \caption{Dynamic Faraday ellipticity in the magnetophotonic crystal. (a) Hysteresis loops under externally-applied magnetic field recorded from the magnetophotonic crystal with $D\_B =227$ nm, at a wavelength of 730 nm demonstrating how the magnitude and sign of the Faraday ellipticity ($\eta\_{F}$) can be controlled through the illumination incidence angle ($\alpha\_{i}$), varying between 0 and 20 degs. (b) The change in Faraday ellipticity ($\delta\eta\_{F}$) measured at different wavelengths. Following the onset of the surface lattice resonance, there is an abrupt change in light ellipticity at various illumination angles (0-20 degrees), which is associated with a maximum in $\delta\eta\_{F}$. The peaks at different incidence angles have been fitted with Lorentzians.
  }
\label{fig3}
\end{figure}

The experimental $\Theta\_F$, $\eta\_F$ and $\theta\_F$ curves all show abrupt features that onset with the excitation of the surface lattice resonance associated with $\lambda\_{R}^{[-1,0]}$ in the $\varphi\_{i}$ = 90$^\circ$ configuration. However, spectrally just prior to this resonance there is the greatest change in MO activity for the smallest change in wavelength. Since this feature is dependent on the spectral position of $\lambda\_{R}^{[-1,0]}$, it can be effectively tuned by varying $\alpha\_{i}$, indicating the potential use of such magnetophotonic crystals as light incidence direction/angular sensors. This is explored in Fig. \ref{fig3}a, where hysteresis loops are recorded through measurements of the transmitted light ellipticity at a wavelength of 730~nm for the nanoantennas with base diameter 227~nm for different $\alpha\_{i}$. The nanoantenna's Tb$_{18}$Co$_{82}$ tops maintain perpendicular magnetic anisotropy even after the lithography process, which is clear from the large remanent magnetization observed in the hysteresis loops in Fig.\ \ref{fig3}a, reducing the magnetic field strength required to saturate the sample along the out-of-plane direction. The dynamic tunability of the MO activity by varying $\alpha\_{i}$ is remarkable in this case, resulting in a dramatic change in the magnitude of $\eta\_F$, where extraordinarily at $\alpha\_{i}$ = 15$^\circ$ the loop is even inverted (see a view of $\eta\_F$ for the spectral region around the surface lattice resonance in Supporting information; it is clear that this sign change in $\eta\_F$ is associated with the migration of the surface lattice resonance to the measurement wavelength of 730~nm).

This is explored further in Fig.\ \ref{fig3}b where the change in Faraday ellipticity ($\delta\eta\_F$) between successive wavelength increments ($\delta\lambda$ = 5~nm) is plotted. Since the gradient of this feature is positive when it coincides with $\lambda\_{R}^{[-1,0]}$ (see inset of Fig.\ \ref{fig3}a), the $\delta\eta\_F$ $<$ 0 data  has been excluded from the fits. It is evident that $\eta\_F$ undergoes a sign change, which in turn is tunable by varying $\alpha\_{i}$. This active tuning modality was previously envisioned for refractive index biochemosensing, where the spectral region of maximum sensitivity can be tuned by varying the angle of incidence, thereby allowing to operate in a spectral region where the analyte solution is minimally absorbing \cite{kazuma2013}. Here we foresee that the deviations from a set angle, i.e., a mechanical tilt, could be employed in high-precision tilt-control systems and detected with high accuracy, simply as reduced MO activity in transmittance. The latter feature starkly differentiates this approach from the currently employed optical systems where reflection is captured by a complex system of mirrors/detectors often with the need of a microelectromechanical (MEMS) array of actuators. Lorentzian functions have been fitted to the $\delta\eta\_F$ data, in order to estimate the spectral width of the abrupt transition in $\eta\_F$. Due to the limited number of data  points on this  abrupt spectral transition, a full estimate of the full-width at half-maximum (FWHM) is difficult to obtain from these fits. However, all values are within the $5-10$ nm range (which is comparable to the wavelength resolution of the setup) with the exception of the $\alpha\_{i}$ = 10$^{\circ}$ where a FWHM of 24 $\pm$ 10~nm is obtained due to the anomalously large error on this particular measurement.

\begin{figure}[t!]
\centering
     \includegraphics[angle = -0,width = \textwidth]{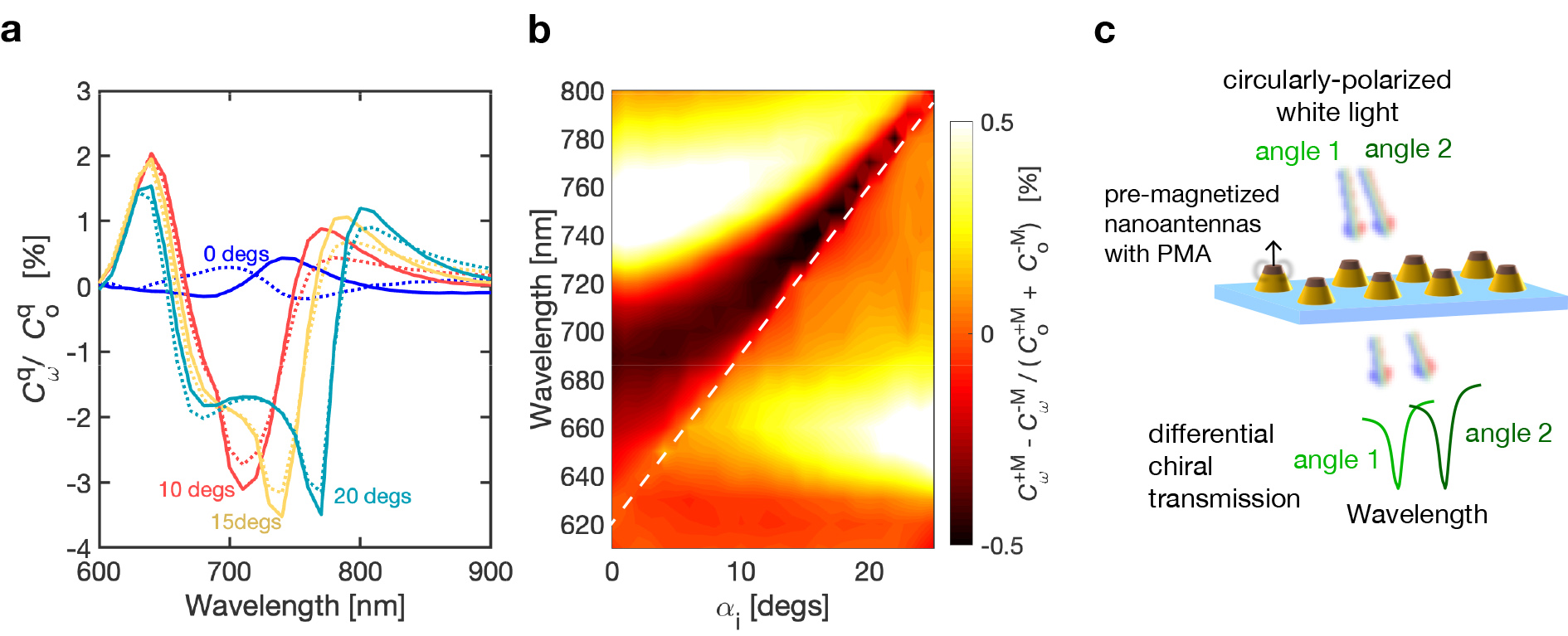}
 \caption[]{Angle-controlled chiral transmittance and mechanical tilt-angle sensing. (a) Spectral dependence of the $C\_{\omega}^{q}/C\_{\circ}$ signals, where $q$ = +$M\_{z}$ or -$M\_{z}$ for the solid and dashed curves respectively. $C\_{\omega}^{q}$ is related to the total circular dichroism for a particular magnetization state, containing both magnetic and non-magnetic contributions. (b) The amplitude of the magnetic modulation of the helicity dependent transmission as function of both wavelength and $\alpha\_{i}$, which relates to the difference between the solid and dashed curves in (a). The dashed white line indicates the expected location of the Rayleigh anomaly calculated from Eq.\ (3). (c) Schematics of the tilt-angle sensing device, where the difference in left- and right- circularly-polarized light, passing through the pre-magnetized magnetophotonic crystal, is detected as spectrally-resolved differential chiral transmission, having sharp spectral dips at distinct wavelengths, depending on the light's angle of incidence.}
\label{fig4}
\end{figure}

Crucially, the perpendicular magnetic anisotropy in this magnetophotonic crystal allows for the measurement of the magnetic differential absorption of circularly polarised light, underpinning $\eta\_F$, without the need for an out-of-plane magnetic field to stabilize the magnetization along the propagation direction of light. When circularly polarized light beam, with a time-varying helicity is incident on  the sample, we can measure the ratio, $C\_{\omega}^{q}/C\_{\circ}^{q}$ which is  proportional to the differential absorption of circularly polarised light (see Methods) for the two opposite polar magnetization states. Here, $C\_{\omega}^{q}$ is the amplitude of the $\omega/2\pi$~=~50~kHz signal from the modulation of the light circular polarization (see Methods), for a fixed polar magnetization $q = \pm  M\_{z}$, while $C\_{\circ}^{q}$ is the DC signal intensity, which contains the helicity independent absorption contribution. Fig.\ \ref{fig4}a shows several spectra for the nanoantennas with base diameter of 227~nm for different values of $\alpha\_{i}$, in the $\varphi\_{i}$ = 90$^{\circ}$ configuration and in zero external magnetic field.  The spectral minima strongly depend on $\alpha\_{i}$.  If we include an external field, the amplitude of $C\_{\omega}^{q}/C\_{\circ}^{q}$ can be further modulated by  reversing the magnetization ($q$ = $+ M\_{z} \rightarrow -M\_{z}$, and vice versa), as indicated by the variation between the dashed and solid curves. The magnetophotonic crystal then exhibits active transmission tunability, whereby absolute transmission can be enhanced or attenuated with the use of a magnetic field. Similar active magnetic transmission tunability has been devised  with magnetoplasmonic chiral nanoantennas \cite{zubritskaya2018}, however, an external field was required to orient the magnetization out-of-plane throughout the measurement, whereas here the external field is only required to set the magnetic  state.  An additional  tuning knob is implemented through the light incidence direction/angle $\alpha\_{i}$, whereby the spectral location of this maximum for magnetic modulation can be tuned with the surface lattice resonance. 

We define a magnetic asymmetry ratio $(C\_{\omega}^{-M\_{z}}-C\_{\omega}^{+M\_{z}}) / (C\_{\circ}^{-M\_{z}}+C\_{\circ}^{+M\_{z}})$, which represents the available helicity-dependent transmission modulation between the two antiparallel magnetization states (see Methods) which is plotted in Fig.\ \ref{fig4}b. The dispersion of the surface lattice resonance calculated from equation (3) is given by the dashed lines. Here, it is clear that the latter dictates the onset wavelength for  the magnetic modulation of the differential circular transmission, meaning that the peak sensitivity can be tuned to arbitrary wavelength between 650~nm - 800~nm. This tunability range is governed by the FWHM of the magnetophotonic crystal transmission spectra.   A maximum magnetic asymmetry ratio of around 0.5\% can be obtained, however, we believe there is enormous scope for improvement through composition optimisation of the RE-TMs and the noble metal thicknesses in the nanoantennas, including exploring new geometries sustaining plasmon optically dark modes, which result in a stronger plasmonic enhancement of the MO activity than achieved with the here-used strongly scattering dipolar plasmons \cite{lopez-ortega2020}. The essential operation of a simple mechanical tilt-control/light incidence optical sensor can be further envisioned as in Fig.\ \ref{fig4}c. The differential chiral transmission ($C\_{\omega}^{q}/C\_{\circ}^{q}$) reports the mechanical tilt/change of light incidence direction angle on the pre-magnetized magnetophotonic crystal by having sharp spectral dips at various wavelengths. We can also envision that by using materials exhibiting all-optical magnetization switching \cite{ciuciulkaite2020} such as the TbCo family of alloys employed here, the need for the external magnetic field to setup the magnetic state of the magnetophotonic crystal or for magnetic transmission modulation can be entirely removed, whereby the transmission would be modulated purely optically at the ultrafast (fs) timescale and allowing for sub-wavelength (nanoscale) miniaturization \cite{alebrand2012,liu2015}.

\section{Conclusion}
In conclusion, our work demonstrates the seamless integration of a rare-earth--transition-metal into magnetophotonic crystals. A strong angular dispersion is engineered through the interference of the Rayleigh anomaly and the nanoantenna's plasmons,  producing a sharp surface lattice resonance in both the optical and MO responses. We showcase dynamic tunability of magnetophotonic crystals using  the light's incidence direction angle, which strongly modifies the MO response, as rationalized using finite-element method electromagnetic simulations. Further, we have shown the magnetic modulation of the differential circular transmission of a magnetophotonic crystal, with measurements performed in zero external magnetic field, exploiting the perpendicular magnetic anisotropy of the magnetic nanoantennas. More generally, the integration of rare-earth--transition-metals within plasmonic nanoantennas offers an exciting platform for highly tunable, ultrafast all-optical switching active magnetophotonic devices \cite{liu2015,maccaferri2020}. Such architectures could also find further application scope where the optical response from magnetophotonic crystals can be tuned by the angle of incidence \cite{haghtalab2020,shi2020} in combination with the reconfigurable magnetic structure \cite{wang2020} steered by all-optical ultrafast magnetization switching or by external magnetic fields.

\section{Experimental Section}

\threesubsection{Sample Fabrication}

The plasmonic nanoantennas are fabricated using a top-down approach, based on the method outlined by Horrer {\it et al}. \cite{horrer2013}. Au(80 nm) films were deposited using electron-beam evaporation onto fused-silica substrates. Later, Al$_{2}$O$_{3}$(3.5 nm)/Tb$_{18}$Co$_{82}$(15 nm)/Al$_{2}$O$_{3}$(2 nm) films were sputter deposited onto these films. The Tb$_{18}$Co$_{82}$ layer was deposited through co-sputtering, with the complete structure being Au(80 nm)/Al$_{2}$O$_{3}$(3.5 nm)/Tb$_{18}$Co$_{82}$(15 nm)/Al$_{2}$O$_{3}$(2 nm). The additional thin Al$_{2}$O$_{3}$ layers were used as capping and isolating layers for the Tb$_{18}$Co$_{82}$. Here, the composition of the film can be varied by adjusted the relative power of the Co and Tb magnetrons. Calibration films were made with different power ratios on the two magnetrons and compositions were verified using Rutherford back scattering. Electron beam lithography was used to define disk shaped apertures in a MicroChem 496PMMA A4 electron-beam resist. Electron-beam evaporation was used to deposit an Al mask through the resist followed by removal of the PMMA mask with Acetone. The resulting structure was then milled at a 5~deg incidence angle with sample rotation, removing all material unprotected by the Al mask. Any remaining Al mask was then removed with the photoresist developer Microdeposit 351, which in this case was used as a selective etcher to target the Al. A conical profile is induced through a combination of the small lateral component of the milling which depends to some extent on the small milling incidence angle \cite{fleischer2010}. In our samples, this results in a constant slope profile of approximately 62~degrees for all nanoantenna arrays.  Therefore, by varying the diameter of the Al mask, the resulting structures can be tuned from truncated to conical profiles.

\threesubsection{Magneto-optical characterisation}

The experimental values of $\theta\_{F}$, $\eta\_{F}$ and $\Theta\_{F}$  were measured using the photoelastic modulator methodology with an applied field of 450~mT  along the light propagation direction, which is described in the Supporting Information. A quadratic polynomial was fitted to the raw $\theta\_{F}$ data in order to subtract the background contribution which arises from the Faraday rotation of the fused-silica substrate, which is strongest for short wavelengths and decreases for longer wavelengths \cite{qiu1998}. For the differential absorption of circularly polarised light measurement, a time varying light polarisation, which alternates between left and right circularly polarised light states at 50 kHz was generated using a photoelastic modulator (PEM) and directed at the sample at normal incidence. This is achieved by passing linearly polarised light orientated at 45$^{\circ}$ to the fast axis of the PEM, with the PEM retardation set to 0.25 wavelengths. Any mechanism in the TNC array which results in a difference in absorption for opposite helicities (including magnetic circular dichroism) will contribute to an oscillating light intensity at the detector at the photoelastic modulator frequency. It is common to express this measurement as the ratio $C\_{\omega}^{q}/C\_{\circ}^{q}$, where $C\_{\omega}^{q}$ is the amplitude of the $\omega$ = 50 kHz signal for a fixed polar magnetization $q = \pm  M\_{z}$, and $C\_{\circ}^{q}$ is the DC signal intensity, which contains the helicity independent absorption contribution. Prior to the measurement, a saturating magnetic field was used to initialise the magnetization along the light propagation direction ($q = +M\_{z}$) and then removed. For the subsequent measurement, the magnetization was saturated in the opposite polar direction ($q = -M\_{z}$) and the measurement repeated.

It is important to note that the spectra in Fig.~\ref{fig4}a contain additional {\it fake} CD contributions, which arise from leaking-in of the large linear dichroism signal as a result of the rectangular array with which the nanostructures are arranged. By observing the difference between the antiparallel magnetization states, these effects, which are independent of the magnetization orientation, can be subtracted out, yielding the available magnetic modulation. We define this magnetic modulation of the helicity dependent transmission as $(C\_{\omega}^{-M\_{z}}-C\_{\omega}^{+M\_{z}}) / (C\_{\circ}^{-M\_{z}}+C\_{\circ}^{+M\_{z}})$, and this quantity is plotted in Fig.~\ref{fig4}b as a function of both $\alpha\_{i}$ and wavelength.

\medskip
\textbf{Supporting Information} \par 
Supporting Information is available from the Wiley Online Library or from the author.

\medskip
\textbf{Acknowledgements} \par 
The authors would like to express their gratitude towards Prof.\ Bengt Lindgren of Uppsala University, Sweden, for fruitful discussions and support with the ellipsometric characterization of TbCo thin film materials. The excellent support and infrastructure of the MyFab facility at the \AA ngstr\"om Laboratory of Uppsala University is also highly appreciated. The authors acknowledge support from the Knut and Alice Wallenberg Foundation project ``{\it Harnessing light and spins through plasmons at the nanoscale}'' (Project No.\ 2015.0060), the Swedish Research Council (Project No.\ 2019-03581), the Swedish Foundation for International Cooperation in Research and Higher Education (Project No.\ KO2016-6889), and the Swedish National Infrastructure for Computing (SNIC). This work is part of a project which has received funding from the European Union's Horizon 2020 research and innovation programme under grant agreement no.\ 737093, ``{\textsc{femtoterabyte}}''. P.V. acknowledges funding from the Spanish Ministry of Science and Innovation under the Maria de Maeztu Units of Excellence Programme (MDM-2016-0618) and the project RTI2018-094881-B-I00 (MICINN/FEDER).

\medskip
\textbf{Author Contributions} \par
R.M.R-R.\ and V.K.\ designed the material and nanofabrication processing, with input from A.D.\ concerning the nanocone design approach. R.M.R-R.\ and A.C.\ carried out the thin film deposition and magnetic characterization. R.M.R-R., A.C., I.-A.C.\ and M.P.\ performed all magneto-optical characterization of the nanoarrays. J.H., R.M.R-R., M.Z., P.V.\ and P.M.O.\ did the electromagnetic modelling and simulations. R.M.R-R.\ and V.K.\ wrote the manuscript with input from J.H., P.V., A.D.\ and P.M.O. All authors discussed the results and commented on the manuscript.


\end{document}